\documentclass{aastex63}
\usepackage{amsmath,bm}

\newcommand{\Alfven}{Alfv\'en }
\newcommand{\Alfvenic}{Alfv\'enic }

\accepted{\today}

\submitjournal{ApJ}

\shorttitle{Anisotropies and scalings}
\shortauthors{Wang et al.}

\graphicspath{{./}{figures/}}

\begin{document}

\title{Observational quantification of three-dimensional anisotropies and scalings of space plasma turbulence at kinetic scales}

\correspondingauthor{Tieyan Wang}
\email{tieyan.wang@gmail.com}

\author[0000-0003-3072-6139]{Tieyan Wang}
\affiliation{RAL Space, Rutherford Appleton Laboratory, Harwell Oxford, Didcot OX11 0QX, UK}

\author{Jiansen He}
\affiliation{School of Earth and Space Sciences, Peking University, Beijing 100871, China}

\author{Olga Alexandrova}
\affiliation{LESIA, Observatoire de Paris, Universit\'e PSL, CNRS, Sorbonne Universit\'e, Univ. Paris Diderot, Sorbonne Paris Cit\'e, 5 place Jules Janssen, 92195 Meudon, France}

\author{Malcolm Dunlop}
\affiliation{School of Space and Environment, Beihang University, Beijing 100191, China}
\affiliation{RAL Space, Rutherford Appleton Laboratory, Harwell Oxford, Didcot OX11 0QX, UK}
\author{Denise Perrone}
\affiliation{ASI - Italian Space Agency, via del Politecnico snc, 00133 Rome, Italy }


\begin{abstract}

A statistical survey of spectral anisotropy of space plasma turbulence is performed using five years measurements from MMS in the magnetosheath. By measuring the five-point second-order structure functions of the magnetic field, we have for the first time quantified the three-dimensional anisotropies and scalings at sub-ion-scales ($<$ 100 km). In the local reference frame $(\hat L_{\perp}, \hat l_{\perp}, \hat l_{\parallel})$ defined with respect to local mean magnetic field $\bm{B}_0$ \citep{2012ApJ...758..120C}, the “statistical eddies” are found to be mostly elongated along $\bm{B}_0$ and shortened in the direction perpendicular to both $\bm{B}_0$ and local field fluctuations. From several $d_i$ (ion inertial length) toward $\sim$ 0.05 $d_i$, the ratio between eddies’ parallel and perpendicular lengths features a trend of rise then fall, whereas the anisotropy in the perpendicular plane appears scale-invariant. Specifically, the anisotropy relations for the total magnetic field at 0.1-1.0 $d_i$ are obtained as $l_{\parallel} \simeq 2.44 \cdot l_{\perp}^{0.71}$, and $L_{\perp} \simeq 1.58 \cdot l_{\perp}^{1.08}$, respectively. Our results provide new observational evidence to compare with phenomenological models and numerical simulations, which may help to better understand the nature of kinetic scale turbulence.

\end{abstract}


\section{Introduction} \label{sec:intro}

The energy distribution at a certain scale (or $\bm{k}$ space) is known to be not isotropic in the turbulence of magnetized plasma, also known as spectral anisotropy \citep{2000ApJ...539..273C}. This particular feature reflect the preferential direction of the energy cascade with respect to the local background magnetic field $\bm{B}_0$ \citep{2009ApJ...698..986P}. Most of our experimental knowledge of space plasma anisotropy comes from in-situ observations made within the solar wind (SW), which is a nearly collisionless plasmas stream released from the Sun \citep{2013LRSP...10....2B}.

At large magnetohydrodynamic (MHD) scales, the pattern of correlation function for the magnetic field at 1 AU has two major components referred to as ``Maltese cross'', exhibiting elongations in both parallel and perpendicular direction with regard to $\bm{B}_0$ \citep{1990JGR....9520673M}. This signature is summarized as the ``slab+2D'' model, which assumes no specific nature of the fluctuations but just describe the fluctuations as a combination of waves with $k_{\parallel}$ and structures with $k_{\perp}$. 
Another type of anisotropy model is based on ``critical-balance (CB)'' conjecture, where the key hypothesis relies on the comparable scale of the linear \Alfven time and turbulence non-linear time in a vanishing cross-helicity system \citep{1995ApJ...438..763G}. As a result, the spectral anisotropy scales as $k_{\parallel} \propto k_{\perp}^{2/3}$. By introducing the idea of ``dynamic alignment'' between magnetic and velocity field fluctuations, \citet{2006PhRvL..96k5002B} modified the non-linear time and established the 3D anisotropic turbulence model, where the eddies have three different coherent scales. Indeed, numerous observations have found agreement between measurements and CB theories \citep{2008PhRvL.101q5005H, 2010ApJ...714L.138L, 2011MNRAS.415.3219C, 2012ApJ...758..120C}. Despite this consistency, recent revisit of anisotropy in the solar wind have reported some puzzling results and raised more concerns to be considered, such as intermittency \citep{2014ApJ...783L...9W, 2016JGRA..121..911P, 2017ApJ...846...49Y}, the discrepancy between velocity field and magnetic field anisotropy \citep{2011PhRvL.106d5001W,2019ApJ...882...21W,2019ApJ...883L...9W,2016ApJ...816L..24Y}, dependence on the heliocentric distance \citep{2013ApJ...773...72H} and solar wind expansion \citep{2018ApJ...853...85V,2019MNRAS.486.3006V}. Moreover, the 3D self-correlation functions are shown to be isotropic in \citep{2019ApJ...871...93W,2019ApJ...882...21W,2019ApJ...883L...9W}. Therefore, the anisotropic nature of the solar wind at MHD scales is still an open question.

At kinetic scales, the turbulence still remains or become much anisotropic (i.e., \citet{2010PhRvL.104y5002C,2015RSPTA.37340152O} and references therein). The standard Kinetic \Alfven Wave (KAW) turbulence model, also on basis of CB conjecture by assuming the linear KAW propagation time to be comparable to nonlinear time, predicts an anisotropy scaling of $k_{\parallel} \propto k_{\perp}^{1/3}$ \citep{2008JGRA..113.5103H,2009ApJS..182..310S}. It is also important to pay attention to the much complicated physical process at kinetic scales than at MHD regime due to plasma kinetic effects (see the review by \citet{2013SSRv..178..101A} and \citet{2020arXiv200401102A}). For example, the modified KAW turbulence model with intermittent 2D structures has $k_{\parallel} \propto k_{\perp}^{2/3}$ \citep{2012ApJ...758L..44B}. \citet{2016JGRA..121....5Z} suggested a model for kinetic-scale \Alfvenic turbulence which incorporate the dispersion and intermittency effects. \citet{2019PhRvR...1a2006B} considered the decisive role played by the tearing instability in setting the aspect ratio of eddies and hence predicted the spectral anisotropy scalings between $k_{\parallel} \lesssim k_{\perp}^{2/3}$ and $k_{\parallel} \lesssim k_{\perp}$. Most recently, \citet{2019arXiv190403903L} proposed a phenomenological model considering the intermittent two-dimensional structures in the plane perpendicular to $\bm{B}_0$. In their model, the prescribed perpendicular aspect ratio of these structures could determine the anisotropy as $k_{\parallel} \propto k_{\perp}^{1/3(\alpha+1)}$, where $\alpha$ is proportional to the space-filling of the turbulence.

In recent high resolution three-dimensional kinetic simulations, the spectral anisotropy has received considerable attentions \citep{ 2018PhRvL.120j5101G,2018ApJ...853...26F, 2019ApJ...879...53A, 2019FrASS...6...64C, 2019arXiv190403903L}. Based on different methods of measuring the anisotropy, dissimilar scaling relations have been found (i.e., $k_{\parallel} \propto k_{\perp}^{1/3}$ in \citet{2018PhRvL.120j5101G} and $k_{\parallel} \propto k_{\perp}$ in \citet{2019ApJ...879...53A}. Specifically, for the analysis based on multi-point local structure functions ($SF$) which will be introduced in Section \ref{sec:data}, the anisotropy tends to become ``frozen'' when approaching ion scales (i.e., $k_{\parallel} \propto k_{\perp}^{0.8}$ in \citet{2019arXiv190403903L}). By comparing the results from three different simulations including the hybrid particle-in-cell (PIC), Eulerian hybrid-Vlasov, and fully kinetic PIC codes, \citet{2019FrASS...6...64C} found that the anisotropy scalings tend to converge to $l_{\parallel} \propto l_{\perp}^{2/3}$ based on a unified analysis of five-point $SF$s.

The Earth's magnetosheath (MSH) offer a unique lab different from SW, such as enhanced compressibility, intermittency, as well as the kinetic instabilities \citep{2008NPGeo..15...95A}. In addition, the spacecraft measurements in MSH tend to cover a wider range of angle between bulk flow velocity and $\bm{B}_0$ as compared with solar wind \citep{2011JGRA..116.6207H}, thus allowing us to diagnose 3D nature of the fluctuations in a relatively short interval. Using Cluster measurements, \citet{2006AnGeo..24.3507M} found the strong anisotropy of the electromagnetic fluctuations, with $k_{\perp}$ extending for two decades within the kinetic range $k d_e \in [0.3, 30]$. \citet{2006PhRvL..96g5002S} showed the anisotropic behaviour up to $k \rho_i =3.5$ in a mirror structure event.
\citet{2008AnGeo..26.3585A} surveyed 6 events and found the dominance of 2D turbulence ($k_{\perp} \gg k_{\parallel}$) above the spectral break in the vicinity of ion scale. In addition, due to the Doppler shift, the magnetic fluctuations have more energy along the $\bm{B} \times \bm{V}$ direction in their analysis. Similar 2D turbulence at kinetic scales was observed in a recent statistical study of magnetic field turbulence in the solar wind \citep{ 2017ApJ...848...45L}. \citet{2011JGRA..116.6207H} computed the spatial correlation functions of both magnetic field and density fluctuations in the 2D $(l_{\parallel}, l_{\perp})$ plane, it is shown that the turbulence close to ion scales is comprised of two populations, where the major component is mostly transverse and the minor one is oblique. Using measurements from \textit{Magnetospheric Multiscale mission} (MMS) \citep{2016SSRv..199....5B}, \citet{2017ApJ...842..122C} studied the two-point $SF$ of the magnetic field in the same plane and provided evidence of strong anisotropy at smaller scales ($11<k \rho_i <57$). Another event recorded by MMS show that parameters such as magnetic field, density, ion velocity, and ion thermal speed all exhibit anisotropy in the spectral index up to $k\rho_i \sim 1$ \citep{2019FrP.....7..184R}.


Despite these case studies, a comprehensive in-situ measurement of the kinetic-scale 3D anisotropy is still lacking. Moreover, to our knowledge, the investigations concerning the scale-dependency of the anisotropy, especially the scaling relations, have not been made yet. The intention of this paper is targeted to address these issues. Based on MMS measurement of magnetic field and ion velocity with unprecedented time-resolution, we applied for the first time five-point second-order $SF$ to statistically quantify the 3D anisotropy of the magnetic turbulence at sub-proton scale ($<$ 100 km). The new observational evidence we obtained, such as the empirical relations of the anisotropy scalings, can be compared with recent theoretical and numerical results, which may facilitate our understanding of kinetic scale turbulence in the space plasmas.


The paper is organized as follows. We describe the data and methods in Section \ref{sec:data}, present an typical event with three-dimensional anisotropy in Section \ref{sec:event}, provide the statistical results in Section \ref{sec:statis}, summarize and discusses our results in Section \ref{sec:conclusion}.

\section{Data and methods} \label{sec:data}
The burst mode data from four MMS spacecraft, including magnetic field (128 Hz) from FIELD instrument \citep{2016SSRv..199..189R} and ion moments (6.7 Hz) from FPI instrument \citep{2016SSRv..199..331P} are used in this study.
349 MSH intervals from September 2015 to December 2019 have been selected for the statistical analysis, whereas a 10 minutes event on 4 October 2017 is presented to show a typical event with anisotropy signatures.

To quantify the anisotropy, we use the five-point second-order $SF$ in this study. Compared with two-point $SF$, five-point $SF$ is more suitable for studying spectral anisotropy at sub-ion regime \citep{2019ApJ...874...75C,2019arXiv190403903L}. More details of the difference between multi-point $SF$s are provided on APPENDIX \ref{app:a1}. 

The five-point $SF$ is the ensemble average of the squared variation $\Delta f$ from 5-point, as function of displacement $\bm{l}$, the $S_2^{(5)}(\bm{l};f)$ is defined as
\begin{equation}
S_2^{(5)}(\bm{l};f)=\langle|\Delta f(\bm{r},\bm{l})|^2\rangle_{\bm{r}}
\end{equation}
The spatial variation from 5-point is measured as  
\begin{equation}
\Delta f(\bm{r},\bm{l})=[f(\bm{r}-2\bm{l})-4f(\bm{r}-\bm{l})+6f(\bm{r})-4f(\bm{r}+\bm{l}))+f(\bm{r}+2\bm{l})]/\sqrt{35}
\end{equation}
Where $\Delta f$ can represent perpendicular, parallel, or total magnetic field fluctuations, $\langle ... \rangle_{\bm{r}}$ is ensemble average over positions $\bm{r}$.

By studying the three-dimensional distribution of the $S_2^{(5)}(\bm{l};f)$ with respect to $\bm{l}$, the “statistical shape” of the eddies in turbulence can be thus inferred from the contours of $S_2^{(5)}(\bm{l};f)$.

In the computation, the velocity field is used to link the time scale with space displacement as $\bm{l}=\tau \cdot \bm{V}$ according to Taylor hypothesis \citep{1938RSPSA.164..476T}, which has been tested in In APPENDIX \ref{app:a2} to be valid for most of our events. Here $\bm{V}$ can be adopted as the mean velocity $\bm{V_{mean}}$ during the interval of interest or considered as the local velocity field $\bm{V_{local}}$, where $\bm{V}$ is interpolated on the resolution of $\bm{B}$ in the calculation. Most of previous works performed in solar wind and magnetosheath use $\bm{V_{mean}}$ for simplicity (i.e., \citet{2010PhRvL.104y5002C, 2012ApJ...758..120C, 2017ApJ...842..122C, 2019ApJ...871...93W, 2019ApJ...882...21W}). Also, the study of electron scale magnetic field structure functions is based on $\bm{V_{mean}}$ \citep{2017ApJ...842..122C}.  In a recent paper by \citet{2018ApJ...853...85V} the authors have considered the effects of local velocity by using $\bm{V_{local}}$ in their analysis of velocity filed structure functions. We use five-point $\bm{V_{local}}$ in this paper, which is defined as $\bm{V}_{local}=[\bm{V}(\bm{r}-2\bm{l})+4\bm{V}(\bm{r}-\bm{l})+6\bm{V}(\bm{r})+4\bm{V}(\bm{r}+\bm{l}))+\bm{V}(\bm{r}+2\bm{l})]/16$.
For the small spatial scales considered here, let us rewrite $\bm{l}=\tau \cdot \bm{V}$ as $\bm{l}=\tau \cdot (\bm{V_0}+\delta\bm{V})$. Since $\tau$ is small, the major contribution of velocity term is from the large scale mean velocity $\bm{V_{0}}$. In other words, the resolution of $\bm{V}$ is not the key factor for interring $\bm{l}$, since it only determines $\delta\bm{V}$, whose amplitude is generally much smaller than $\bm{V_{0}}$ in the magnetosheath environment under this study. To verify this point, we have performed and compared the analyses on basis of $\bm{V_{mean}}$ and $\bm{V_{local}}$, and the results turn out to be nearly identical. Hence, we justify that the effects of interpolating $\bm{V}$ at the time points of $\bm{B}$, or only considering $\bm{V_{mean}}$ is negligible in the analysis of small-scale magnetic field structure functions.

Once the $SF$ with respect to $\bm{l}$ is computed, it can be projected into local coordinates with respect to the local magnetic field $\bm{B}_{local}$ as in \citet{2012ApJ...758..120C,2018ApJ...853...85V} to study its 3D features. This coorinates allows us to compare the results with recent simulations on basis of the same reference frame \citep{2019arXiv190403903L, 2019FrASS...6...64C}, and is consistent with previous choice of studying spectral anisotropy at various scales \citep{2012ApJ...758..120C, 2017ApJ...842..122C, 2018ApJ...853...85V, 2019ApJ...882...21W}. In the Cartesian coordinates system $(\hat L_{\perp},\hat l_{\perp},\hat l_{\parallel})$, $\hat l_{\parallel}$ is along $\bm{B}_{local}=[\bm{B}(\bm{r}-2\bm{l})+4\bm{B}(\bm{r}-\bm{l})+6\bm{B}(\bm{r})+4\bm{B}(\bm{r}+\bm{l}))+\bm{B}(\bm{r}+2\bm{l})]/16$, $\hat L_{\perp}$ is the ''displacement direction'' along $\delta \bm B_{\perp}=\bm{B}_{local} \times [\delta \bm{B} \times \bm{B}_{local}]$, and $\hat l_{\perp}=\hat l_{\parallel} \times \hat L_{\perp}$. The Cartesian system can be also converted into spherical polar coordinates system as $(l,\theta_{B},\phi_{\delta B_{\perp}})$, where $\theta_{B}$ represent the angle between $\bm{B}_{local}$ and $\bm{l}$, $\phi_{\delta B_{\perp}}$ represent the angle between $L_{\perp}$ and the projection of $\bm{l}$ on the plane perpendicular to $\bm{B}_{local}$. Similar to \citet{2012ApJ...758..120C}, angles greater than 90$^{\circ}$ are reflected below 90$^{\circ}$ to improve scaling measurements accuracy. Specifically, by setting the ranges of $\theta_{B}$ and $\phi_{\delta B_{\perp}}$, the $SF$ in the three orthogonal directions can be obtained as 
\begin{align}
 S_2^{(5)}(L_{\perp};f)\equiv S_2^{(5)}(L_{\perp};f,\,85^{\circ}<\theta_{B}<90^{\circ}, 0^{\circ}<\phi_{\delta B_{\perp}}<5^{\circ}) \\
  S_2^{(5)}(l_{\perp};f)\equiv S_2^{(5)}(l_{\perp};f,\,85^{\circ}<\theta_{B}<90^{\circ}, 85^{\circ}<\phi_{\delta B_{\perp}}<90^{\circ}) \\
  S_2^{(5)}(l_{\parallel};f)\equiv S_2^{(5)}(l_{\parallel};f,\,0^{\circ}<\theta_{B}<5^{\circ}, 0^{\circ}<\phi_{\delta B_{\perp}}<90^{\circ})
\end{align}
By equating the value between pairs of $S_2(l_{\perp})$, $S_2(l_{\parallel})$, and $S_2(L_{\perp})$, we could infer the anisotropy relation between $l_{\perp}$, $l_{\parallel}$, and $L_{\perp}$.

\section{example of local 3D turbulence} \label{sec:event}

\begin{figure}[ht!]
\plotone{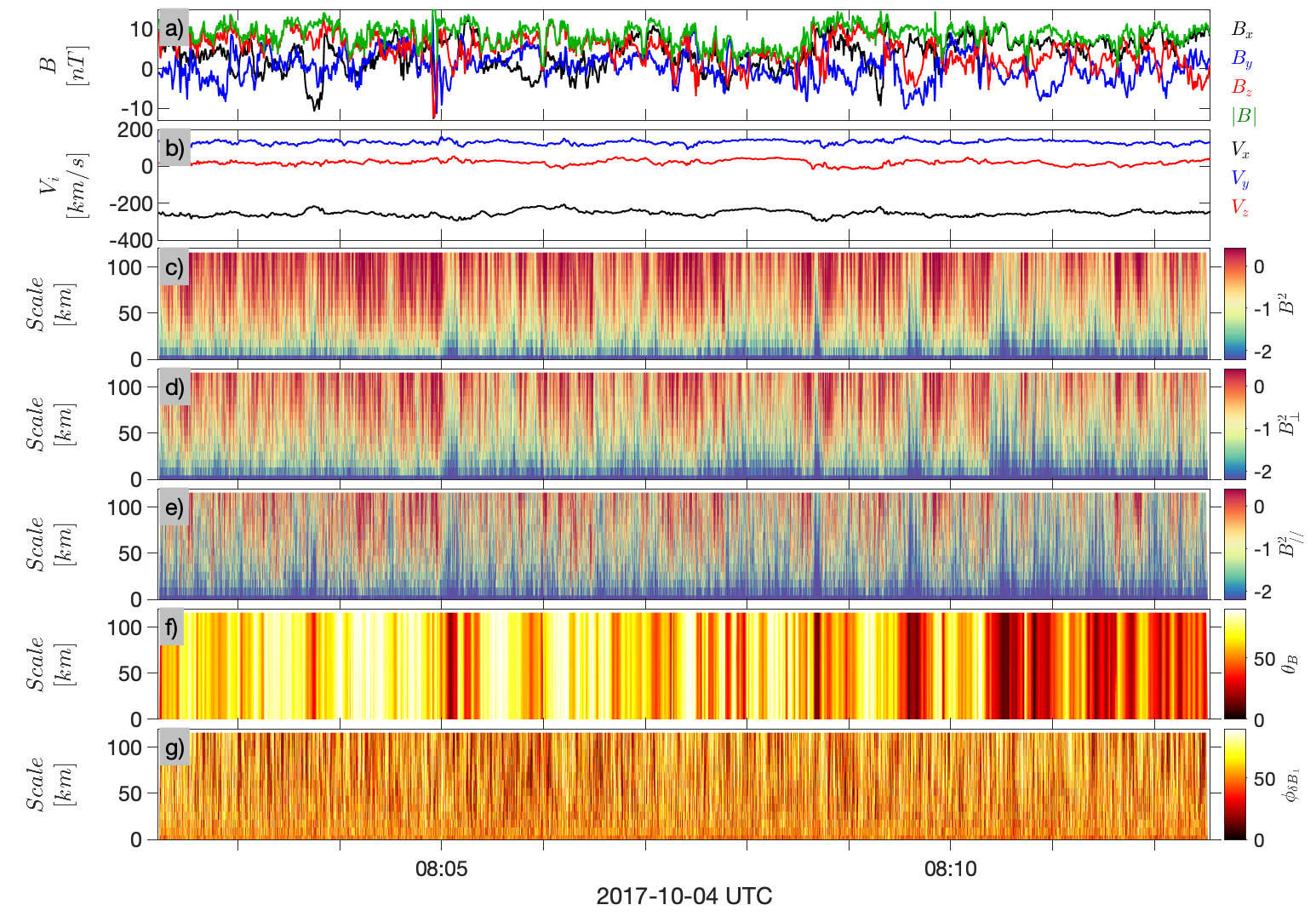}
\caption{Event overview. a) Three components and the strength of the magnetic field. b) Three components of velocity field. c-e) Instantaneous total, perpendicular, and parallel magnetic energy as a function of scale and time. f) Instantaneous angle between local magnetic field and space displacement vector $\bm{l}$, as a function of scale and time. g) Instantaneous angle between $L_{\perp}$ and the projection of $\bm{l}$ on the plane perpendicular to local magnetic field.
\label{fig:Event_OV}}
\end{figure}

Here we present an example with clear signatures of 3D anisotropy. The event is observed downstream of the quasi-parallel shock during 08:02:13-08:12:33 on 4 October 2017. Figure \ref{fig:Event_OV} shows the overview of the event. As plotted in Figure \ref{fig:Event_OV}a-b, the magnetic field is around 6.08 (0.61, 0.06, 0.79) $\pm$ (4.1, 3.6, 3.8) nT, exhibiting numerous large directional changes while keeping its magnitude. In contrast, the ion velocity is quite stable at 288 (-0.88, 0.46, 0.07) $\pm$ (15.5, 10.2, 13.5) km/s. Figure \ref{fig:Event_OV}c shows the instantaneous increment of the total magnetic energy $\delta B^2$ as a function of scale and time. The magnitude of $\delta B^2$ generally increase with the increase of scales and is changing intermittently with time. Similar to $\delta B^2$, the instantaneous increments of the perpendicular energy $\delta B_{\perp}^2$ and parallel energy $\delta B_{\parallel}^2$ also exhibit the same trend with respect to spatial scale as seen in Figure \ref{fig:Event_OV}d-e, while the former one is stronger. Figure \ref{fig:Event_OV}f-g plots the corresponding $\theta_{B}$ and $\phi_{\delta B_{\perp}}$, respectively. Due to the rapid rotations of magnetic field, the distribution of $\theta_{B}$ and $\phi_{\delta B_{\perp}}$ covers a wide range within $(0,\pi)$ during the whole interval, thus allowing us to infer the 3D anisotropy of magnetic turbulence with sufficient data points.

\begin{figure}[ht!]
\centerline{\includegraphics[width=1\textwidth,trim={0cm 0cm 0cm 0cm},clip=]{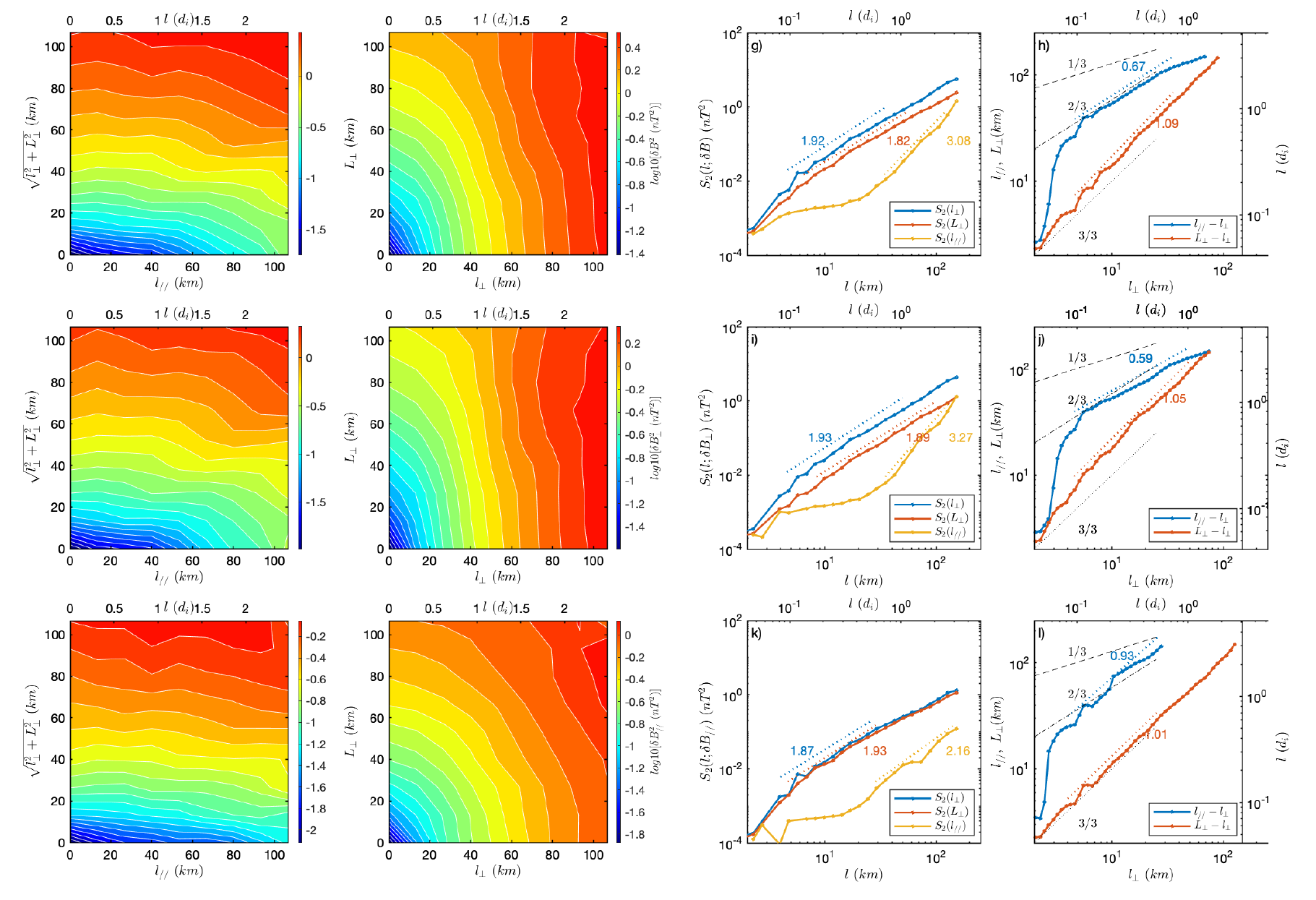}}
\caption{The $SF$s in the 2D plane, together with the 1D $SF$s and their anisotropy scalings. The left two columns include: (1) 2D $SF$s as a function of $(l_{\parallel},\sqrt{l_{\perp}^2+L_{\perp}^2})$, and (2)$SF$s as a function of $(l_{\perp}, L_{\perp})$. The right two columns plot respectively: (1) 1D $SF$ as a function of $l_{\perp}$, $L_{\perp}$, and $l_{\parallel}$, and the relations between $l_{\parallel}$, $l_{\perp}$, and $L_{\perp}$. For each panel, the first, second, and third rows represent $SF$s of the total, perpendicular, and parallel magnetic field, respectively.}
\label{fig:SF_ov}
\end{figure}

Figure \ref{fig:SF_ov} present the structure functions and the corresponding anisotropy scalings for the above event. The values of $SF$s are obtained from four MMS spacecraft, then binned and averaged. Each bin is required to have a minimum number of 200 data points to ensure reliable results as in \citet{2010PhRvL.104y5002C}. For the $SF$ of the total magnetic field energy as projected in the $(l_{\parallel},\sqrt{l_{\perp}^2+L_{\perp}^2})$ plane, the contours of $S_2^{(5)} (\bm{l};\bm{B})$ are elongated in the parallel directions (Figure \ref{fig:SF_ov}a), where the values at perpendicular direction are much larger than the ones in the parallel direction (i.e. $S_2^{(5)} (\bm{l};\bm{B})$ at $l_{\perp}$ = 60 km is more than 100 times larger than the one at $l_{\parallel}$ = 60 km). This signature indicate sub-ion-scale ($l<$ 2 $d_i$) anisotropy with $k_{\perp} \gg k_{\parallel}$, and is in agreement with \citet{2017ApJ...842..122C}. Furthermore, we find that the contours of $S_2^{(5)} (\bm{l};\bm{B})$ are also elongated in the ``displacement'' direction as seen in the $(l_{\perp}, L_{\perp})$ plane (Figure \ref{fig:SF_ov}b), suggesting the three-dimensional characteristics of the anisotropy. In addition to $S_2^{(5)} (\bm{l};\bm{B})$, we also consider the contribution of $SF$ by the perpendicular and parallel magnetic field, namely the $S_2^{(5)} (\bm{l};B_{\perp})$ and $S_2^{(5)} (\bm{l};B_{\parallel})$. As seen in Figure \ref{fig:SF_ov}c-d, the pattern for the contours of $S_2^{(5)}  (\bm{l};B_{\perp})$ is nearly comparable to the ones of $S_2^{(5)}  (\bm{l};\bm{B})$ in Figure \ref{fig:SF_ov}a-b. But the contours in Figure \ref{fig:SF_ov}e are flatter than the ones in Figure \ref{fig:SF_ov}c, meaning the anisotropy of $S_2^{(5)}  (\bm{l};B_{\parallel})$ is slightly stronger than $S_2^{(5)} (\bm{l};B_{\perp})$ in the $(l_{\parallel},\sqrt{l_{\perp}^2+L_{\perp}^2})$ plane. The much-elongated compressive fluctuations along $\bm{B}_0$ is consistent with solar wind observations in \citet{2010PhRvL.104y5002C, 2011MNRAS.415.3219C, 2012ApJ...758..120C}, which may imply the less damped state of the more anisotropic fluctuations. On the contrary, the contours of $S_2^{(5)}  (\bm{l};B_{\parallel})$ are roughly isotropic in the $(l_{\perp}, L_{\perp})$ plane, meaning the absence of gradient in the perpendicular plane for the parallels fluctuations.

Let us draw the attention of the reader to the point that, the sampling of $SFs$ in the perpendicular and parallel direction is dissimilar as seen in Figure \ref{fig:Event_OV}f, whereas the parallel data are much discretely distributed. To test whether the stationarity of the sampling will have an effect on the results of $SFs$, we have divided the time-series into two sub intervals and analyse the $SFs$ separately. During Interval 1 (08: 02: 13 - 08: 07: 13 UT), the measurements along parallel directions (i.e., $\theta_B<$5$^{\circ}$) are less than the ones at oblique directions ($\theta_B>$50$^{\circ}$). In contrast, during Interval 2 (08: 07: 13 - 08: 12: 33 UT), the measurements along parallel directions are much frequent and the overall sampling are more homogeneous than Interval 1. As expected, the $SFs$ for Interval 1 have no measurements within the range of 60 km $<l_{\parallel}<$ 100 km, 0 km $<\sqrt{l_{\perp}^2+L_{\perp}^2}<$18 km), while the SFs for Interval 2 cover the complete wavenumber space. More importantly, the extension feature of the contours along $l_{\parallel}$ direction in these two sub-intervals appears quite similar as compared with the results from the whole interval. Hence, the anisotropic features of the turbulence could be viewed as stationary regardless of the interval selection. In fact, as the first step of computing $SFs$, we calculate the time difference between continuous sampling points rather than discontinuous sampling points. As the second step, the calculated time differences satisfying certain $\theta_B$ conditions are collected together from discretely(discontinuously) distributed time points. This discontinuous collection will not significantly influence the analysis results as long as the whole time interval is statistically time stationary.

To inspect the scale-dependency of the anisotropy more precisely, we have computed the $SF$ in the three orthogonal directions as defined by equation (3)-(5). 
For the 1D $SF$ of the total magnetic field shown in Figure \ref{fig:SF_ov}g, the relation of $S_2(l_{\perp})> S_2(L_{\perp})> S_2(l_{\parallel})$ are satisfied at all scales as expected, thus confirming the 3D nature of anisotropy again. Moreover, this anisotropy is found to be scale-dependent. For example, at energy level of 0.01 nT$^2$, the perpendicular length of the ``statistical eddy”, $l_{\perp}$ $\sim$ 6 km is smaller than the ``displacement” length $L_{\perp}$ $\sim$ 8 km, while the parallel length is much larger at $l_{\parallel}$ $\sim$ 35 km. As the energy level increase to 1 nT$^2$, $l_{\perp}$, $L_{\perp}$, and $l_{\parallel}$ becomes approximately 60 km, 90 km, 150 km, respectively. The change of $l_{\perp}: L_{\perp}: l_{\parallel}$ ratio from 0.17: 0.23: 1 to 0.4: 0.6: 1 suggests that as scales increase, the anisotropy between parallel and perpendicular lengths becomes weak, while the anisotropy between two perpendicular lengths in the perpendicular plane remains almost unchanged. Setting an energy range as [0.01, 0.5] nT$^2$, the $SF$s could be fitted by the power laws as $l_{\perp}^{1.92}$, $L_{\perp}^{1.82}$, and $l_{\parallel}^{3.08}$, where the standard error of the mean is 0.09, 0.08, 0.10, respectively. The power law index of the second-order structure function, $g$, is usually related to the power spectral index, $\alpha$, by $\alpha=g+1$ \citep{2010PhRvL.104y5002C}. Hence the spectral indices in the three directions are 2.92, 2.82, and 4.08, respectively. The perpendicular spectral indicies are close to 8/3 but steeper than 7/3, which are consistent with previous findings both in the MSH and SW \citep{2008AnGeo..26.3585A, 2014ApJ...789L..28H, 2017MNRAS.466..945M, 2017ApJ...842..122C, 2009PhRvL.103p5003A}. For the $SF$s of $\delta B_{\perp}$, the trends plotted in Figure \ref{fig:SF_ov}i are essentially the same compared with the results of $\delta B$ in Figure \ref{fig:SF_ov}g, suggesting a dominant contribution of perpendicular magnetic field fluctuations to $SF$s. This point agrees with the ``variance anisotropy” found in \citet{2010PhRvL.104y5002C} and is again supported by examining $SF$s of $\delta B_{\parallel}$ in Figure \ref{fig:SF_ov}k, whose magnitudes are weaker than $SF$s of $\delta B_{\perp}$ in Figure \ref{fig:SF_ov}g.

Figure \ref{fig:SF_ov}h displays the anisotropy relations for $\delta B$, where the blue curve represent $l_{\parallel}$ vs $l_{\perp}$ and the red curve represent $L_{\perp}$ vs $l_{\perp}$. On one hand, as the perpendicular scales decrease from 1.0 $d_i$ to under 0.05 $d_i$, the ratio of $l_{\parallel}/l_{\perp}$ first increase and reached a maximum of $\sim$ 8 at $\sim$ 0.1 $d_i$, whereas the anisotropy scaling obeys $l_{\parallel} \propto l_{\perp}^{0.67}$. Then the ratio of $l_{\parallel}/l_{\perp}$ begins to decrease and finally approaches 1 at $\sim$ 0.04 $d_i$. On the other hand, the ratio of $L_{\perp}/l_{\perp}$ keeps steady around 1.5, obeying $L_{\perp} \propto l_{\perp}^{1.09}$. As expected, the anisotropy relations for $\delta B_{\perp}$ as shown in Figure \ref{fig:SF_ov}j is quite similar with the relations in Figure \ref{fig:SF_ov}h, following $l_{\parallel} \propto l_{\perp}^{0.59}$ and $L_{\perp} \propto l_{\perp}^{1.05}$. Nevertheless, the anisotropy relations for $\delta B_{\parallel}$ as shown in Figure \ref{fig:SF_ov}l are dissimilar, following $l_{\parallel} \propto l_{\perp}^{0.93}$ and $L_{\perp} \propto l_{\perp}^{1.01}$.

\section{Statistical analysis of the anisotropy scalings} \label{sec:statis}

\begin{figure}[ht!]
\centerline{\includegraphics[width=1\textwidth,trim={0cm 0cm 0cm 0cm},clip=]{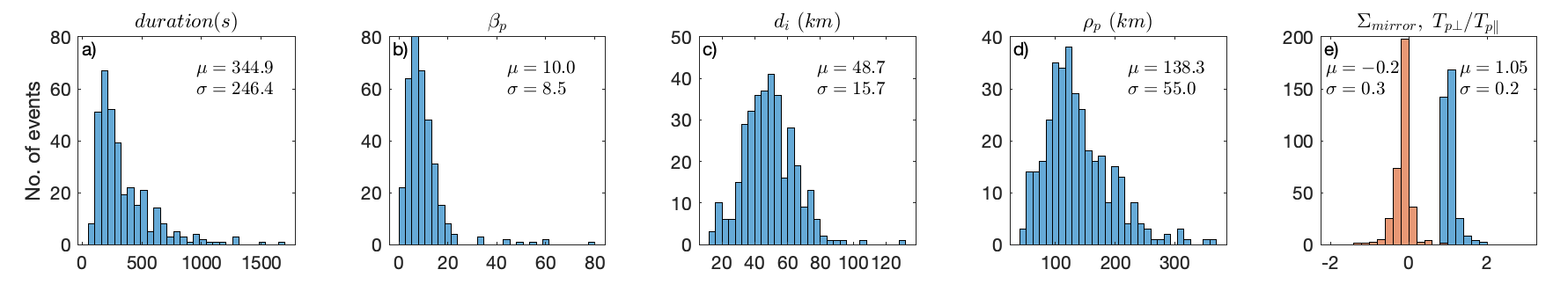}}
\caption{Histograms of the events duration and plasma characteristic parameters.}
\label{fig:STA_par}
\end{figure}

In this section, the sub-ion-scale anisotropy relations are investigated comprehensively based on a statistically survey of 349 intervals during 2015-2019, when MMS instruments was in burst mode. These intervals are tagged as ``magneotosheath” on the MMS science data center website. In addition, to avoid the influence of shock or magnetopuase, the ion and electron energy spectrogram have been checked by eye to make sure they exhibit typical broadband MSH signatures and are time stationary. 
As a result, 349 intervals with an average duration of $\sim$ 5.8 minutes have been selected. Figure \ref{fig:STA_par} shows the histograms for the events duration and plasma parameters, together with their mean value and standard deviations. The events duration, proton beta $\beta_p$, ion inertial length $d_i$, and proton gyroradius $\rho_{p}$, cover the range of [60, 1680] s, [0.3, 80], [15, 130] km, and [45, 370] km respectively. The mean value of temperature anisotropy $T_{p \perp}/T_{p \parallel}$ is 1.05. The distribution of mirror mode threshold $\Sigma_{mirror}=T_{p \perp}/T_{p \parallel}-1/\beta_{p\perp}-1$, is also shown in Figure \ref{fig:STA_par}e. Since most of the values are negative, the influence of mirror instability is not strong in our database.

\begin{figure}[ht!]
\centerline{\includegraphics[width=1\textwidth,trim={0cm 0cm 0cm 0cm},clip=]{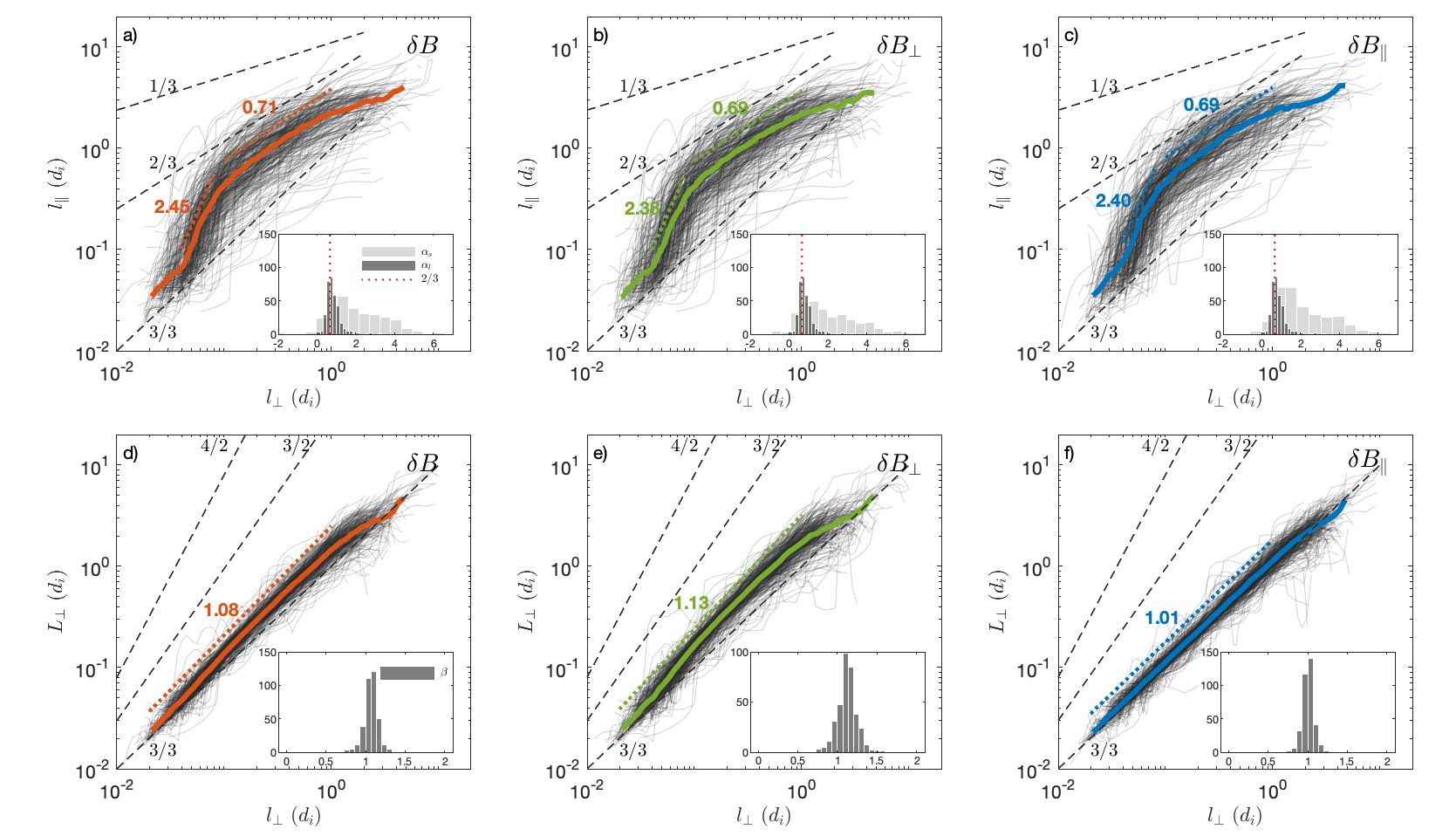}}
\caption{Statistical analysis of the 3D anisotropy scalings. The first, second, and third column shows respectively the anisotropy of $\delta B$, $\delta B_{\perp}$, and $\delta B_{\parallel}$. The light grey curves represent the superposition of the statistical results, the median value and fitted results are overplotted in bold solid and dotted lines. The dashed lines represent specifically, the reference scalings with slope of 1/3, 2/3, 1 at top panels, and 4/2, 3/2, 1 at bottom panels. The histogram of the anisotropy scalings at 0.04 $d_i$ -- 0.08 $d_i$ (light grey), and 0.1 $d_i$ -- 1.0 $d_i$ (dark grey) are inserted in top panels. Similarly, the histograms of the anisotropy at 0.02 $d_i$ -- 1.0 $d_i$ (light grey) are inserted in bottom panels.}
\label{fig:SF_sta}
\end{figure}

Figure \ref{fig:SF_sta} presents the statistical results of the anisotropy relations. Concerning the parallel$-$perpendicular anisotropy of the total magnetic field $\delta B$, as revealed by the unique feature of the superimposed results in Figure \ref{fig:SF_sta}a, a large proportion of events exhibit analogous trend. For the median value of the data, when the scales decrease from 10 $d_i$ to 0.01 $d_i$, the anisotropy level as reflected from the vertical deviation from the isotropy reference line, displays a trend of rise then fall, with the break point occurring near 0.1 $d_i$. The fit of an empirical anisotropy relation $l_{\parallel}=a_0 \cdot l_{\perp}^{\alpha}$ yield $a_0$ of 2.44 and a scaling of $\alpha_l=0.71\pm0.03$ at large scales within [0.1, 1] $d_i$. Compared with three reference scaling-laws of $\alpha$ = $1/3$, $2/3$, and $3/3$, the fitted scaling is closer to $2/3$. At smaller scales within [0.04, 0.08] $d_i$, we find $a_0=170$, and $\alpha_s=2.45\pm0.35$. Likewise, the anisotropy relations of $\delta B_{\perp}$ and $\delta B_{\parallel}$ in Figure \ref{fig:SF_sta}b-c display similar trends as Figure \ref{fig:SF_sta}a, where the anisotropy scalings at [0.1, 1.0] $d_i$ are obtained as $\alpha_l=0.69\pm0.03$ and $\alpha_l=0.69\pm0.04$ and the scalings at [0.04, 0.08] $d_i$ are $\alpha_s =2.38\pm0.32$ and $\alpha_s =2.40\pm0.14$, respectively. By examining the distribution of power-law scalings for each individual event as shown in the histograms in Figure \ref{fig:SF_sta}, we confirm that for the scalings at [0.1, 1.0] $d_i$ (dark grey), a summary of over 100 events have a scaling centered near $2/3$ (red dotted lines). However, as shown in the light grey histogram of the scalings at [0.04, 0.08] $d_i$, the scalings are broadly distributed within [0, 4].

Regarding the anisotropy of $\delta B$ and $\delta B_{\perp}$ in the perpendicular plane, Figure \ref{fig:SF_sta}d-e show that for most of the data, although $L_{\perp}>l_{\perp}$, the anisotropy level is stable since the scalings are close to 1. In addition, the empirical relation $L_{\perp}=b_0 \cdot l_{\perp}^{\beta}$ for the median value are fitted as $L_{\perp}=1.58 \cdot l_{\perp}^{1.08\pm0.01}$, $L_{\perp}=2.03\cdot l_{\perp}^{1.13\pm0.02}$ at [0.02, 1.0] $d_i$, respectively. Lastly, the anisotropy for $\delta B_{\parallel}$ vanishes and follows a nearly isotropic relation of $L_{\perp}=1.15 \cdot l_{\perp}^{1.01\pm0.003}$.

\section{Conclusion and Discussion} \label{sec:conclusion}

In this paper, we have conducted a statistical survey of the sub-ion-scale anisotropy of the turbulence in the Earth’s magnetosheath. By measuring the five-point second-order $SF$s of the magnetic field, the three-dimensional structures of the turbulence have been quantitatively characterized. Specifically, the three characteristic lengths of the eddies are found to roughly satisfy $l_{\parallel} > L_{\perp} > l_{\perp}$ in the local reference frame defined by \citet{2012ApJ...758..120C}. As for the scale-dependency of the anisotropy inferred from $SF$s of the total magnetic field, (1) the parallel$-$perpendicular anisotropy as revealed by the ratio of $l_{\parallel}/ l_{\perp}$, shows an increase trend towards small scales and obeys a scaling of $l_{\parallel} \simeq 2.44 \cdot l_{\perp}^{0.71}$ between 0.1 $d_i$ and 1 $d_i$, then it decreases and obeys $l_{\parallel} \simeq 170 \cdot l_{\perp}^{2.45}$ between 0.04 $d_i$ and 0.08 $d_i$. (2) the anisotropy in the perpendicular plane as revealed by the ratio of $L_{\perp}/ l_{\perp}$, is generally weaker than the ratio of $ l_{\parallel}/ l_{\perp}$. Moreover, this anisotropy is scale-invariant, displaying a scaling of $L_{\perp} \simeq 1.58 \cdot l_{\perp}^{1.08}$.
 
Interestingly, the parallel$-$perpendicular anisotropy tends to become increasingly isotropic when approaching both large scales $\sim$ 4 $d_i$ and small scales $\sim$ 0.04 $d_i$ (Figure \ref{fig:SF_sta} a-c). This large-scale isotropy may reflect similar structures as in the isotropic solar wind reported recently (i.e.,\citet{2019ApJ...871...93W, 2019ApJ...882...21W, 2019ApJ...883L...9W}), while the small-scale isotropy has not been reported before to our knowledge. Possible explanations for such isotropy include the weakening of perpendicular cascade and the influence of ion cyclotron waves. Indeed, there are a few events with $l_{\perp}> l_{\parallel}$ in the database, where the existence of ICW has been confirmed by checking the polarisation state of the fluctuations. In a few other events, we also find coexistence of ICW and 2D $l_{\parallel}> l_{\perp}$ structures through inspecting the $SF$s in the 2D $(l_{\parallel}, l_{\perp})$ plane.

The scaling of $l_{\parallel} \propto l_{\perp}^{0.71}$ observed in this work is different from traditional KAW theory of $l_{\parallel} \propto l_{\perp}^{1/3}$, but is close to the theoretical prediction of $l_{\parallel} \propto l_{\perp}^{2/3}$ from \citet{2012ApJ...758L..44B}, \citet{2019PhRvR...1a2006B} and simulation from \citet{2019FrASS...6...64C}. It also corresponds to $\alpha=1.13$ in the framework of the model by \citet{2019arXiv190403903L}. Hence a modified CB premise may be needed to understand the kinetic turbulence in magnetosheath. We note that, for most of the events, the week anisotropy in the perpendicular plane is inconsistent with results of \citep{2019PhRvR...1a2006B}, which predicts a much stronger anisotropy and a steeper scaling of $L_{\perp} \propto l_{\perp}^{3/2}$ or $L_{\perp} \propto l_{\perp}^{2}$, depending on different current sheet configurations for the tearing instability. Capturing the active signatures of current sheet disruption$/$reconnection (i.e., \citep{2017MNRAS.468.4862M, 2017PhRvL.118x5101L}) from in-situ observation is a challenging task, but will contribute to understand its effects on the anisotropy.

\begin{figure}[ht!]
\centerline{\includegraphics[width=1\textwidth,trim={0cm 0cm 0cm 0cm},clip=]{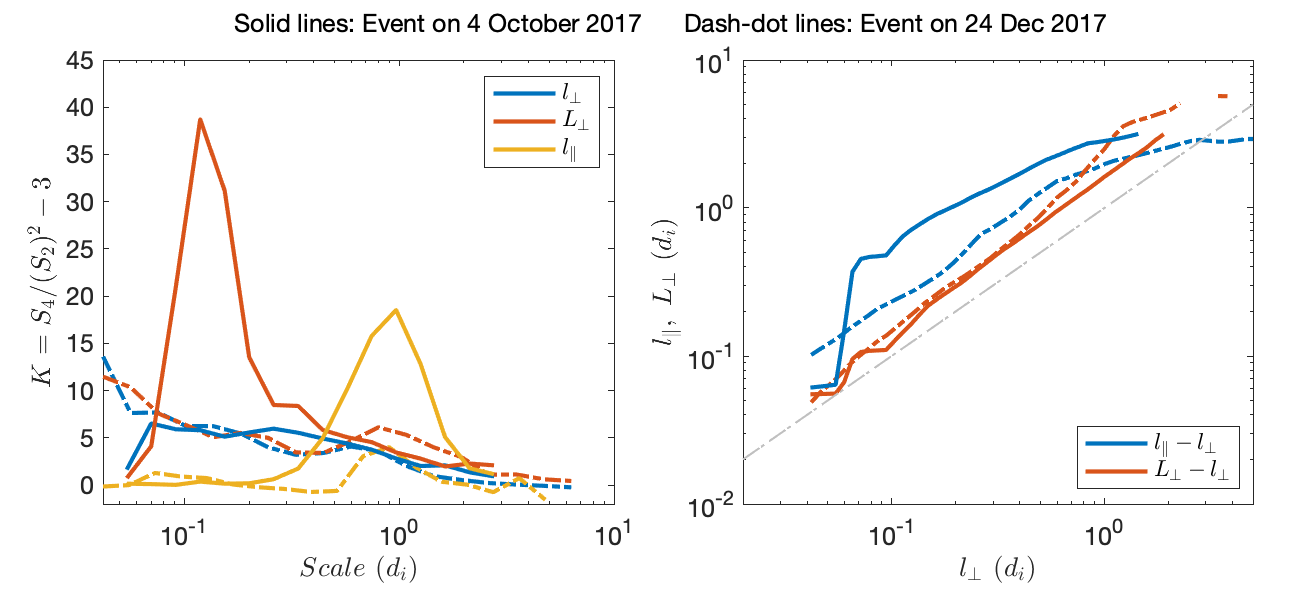}}
\caption{Comparisons of the intermittency and anisotropy relation between two events during 08: 02: 13 - 08: 12: 33 on 4 October 2017 and 01: 04: 43 - 01: 11: 53 on 24 December 2017.}
\label{fig:Intermittency}
\end{figure}

The spectral anisotropy of kinetic plasma turbulence is believed to be associated with dispersion and intermittency effects \citep{2016JGRA..121....5Z, 2019arXiv190403903L}. For example, the anisotropic scalings are different below and above ion cyclotron frequency and also differs for sheet and tube like turbulence \citep{2016JGRA..121....5Z}. 
To illustrate the possible connection between intermittency and spectral anisotropy, we specifically compare the results from two events. Figure \ref{fig:Intermittency} shows the excess Kurtosis and anisotropy relation for the total magnetic field. The excess Kurtosis is defined as $K(l)=S_{4} (l)/S_2 (l)^2-3$, where $S_4 (l)$ is the fourth-order structure function. In both panels of Figure \ref{fig:Intermittency}, the solid lines represent the results from event 1, which is recorded on 4 Oct 2017 and is used as our example of spectral anisotropy in section 3, while the dash-dot lines represent the results from event 2, which is recorded on 24 Dec 2017. As plotted in three directions $l_{\perp}$, $L_{\perp}$, $l_{\parallel}$, the value of $K(l)$ is around zero at large scales, meaning the roughly Gaussian distribution of the fluctuations. Toward small scales, $K(l)$ displays an increase tendency before it drops down. Such non-Gaussian statistics ($K(l)>0$) confirms the presence of intermittency in the magnetosheath, while the scale-dependent profile of Kurtosis is similar with solar wind observations \citep{2019ApJ...873...80H}. For the kinetic scale parallel-perpendicular anisotropy, we find that it can be considerably affected by the intermittency. As shown in Figure \ref{fig:Intermittency}, the stronger the intermittency (see the larger Kurtosis of the solid lines in the left panel), the stronger the anisotropy level (see the larger vertical distances between the solid blue curve and the grey line in the right panel). This phenomenon is consistent with previously observations at large scales, which emphasize the key role of intermittency in generating the spectral anisotropy (i.e., \citep{2014ApJ...783L...9W, 2016JGRA..121..911P, 2018ApJ...855...69Y}). For the anisotropy in the perpendicular plane, the anisotropy levels (see the two red curves in the right panel) of these two events are much smaller as compared with the parallel-perpendicular anisotropies. The reason for such weak anisotropy still remains to be explored. We note that it has been proposed that the axisymmetric, 2D ($k_{\perp}>k_{\parallel}$ fluctuations can be observed with non-axisymmetric features in the spacecraft frame due to a sampling effect (i.e., \citet{2008AnGeo..26.3585A, 2011PhRvL.107i5002T, 2017ApJ...848...45L} and Matteini et al. 2020, submitted). Therefore, the spectral anisotropy (non-axisymmetric) in the perpendicular plane needs be cautiously interpreted with such effects being quantitively explored in the future. Lastly, we note that both of the events still have non-Gaussian fluctuations (as seen in the non-zero values of the Kurtosis), hence the absence of isotropic $l_{\parallel} \approx L_{\perp} \approx l_{\perp}$ relation is not contradictory to previous studies, which found isotropy when the intermittency is removed (i.e., \citep{2016JGRA..121..911P}). We plan to conduct a much comprehensive analysis to understand how intermittency influence spectral anisotropy in a future work, particularly focusing on comparing the role of different coherent structures on the anisotropy (i.e., 2D tube-like vortices in \citet{2019ApJ...871L..22W} or 1D current sheets.

\acknowledgments
We greatly appreciate the MMS development and operations teams, as well as the instrument PIs, for data access and support. This work was supported by the Marie Skłodowska-Curie grant No. 665593 from the European Union’s Horizon 2020 research and innovation programme. J.-S. He is supported by NSFC under 41874200 and 41421003. 

\vspace{5mm}

\appendix
\section{Comparison of multi-point structure functions}\label{app:a1}

\begin{figure}[ht!]
\centerline{\includegraphics[width=1\textwidth,trim={0cm 0cm 0cm 0cm},clip=]{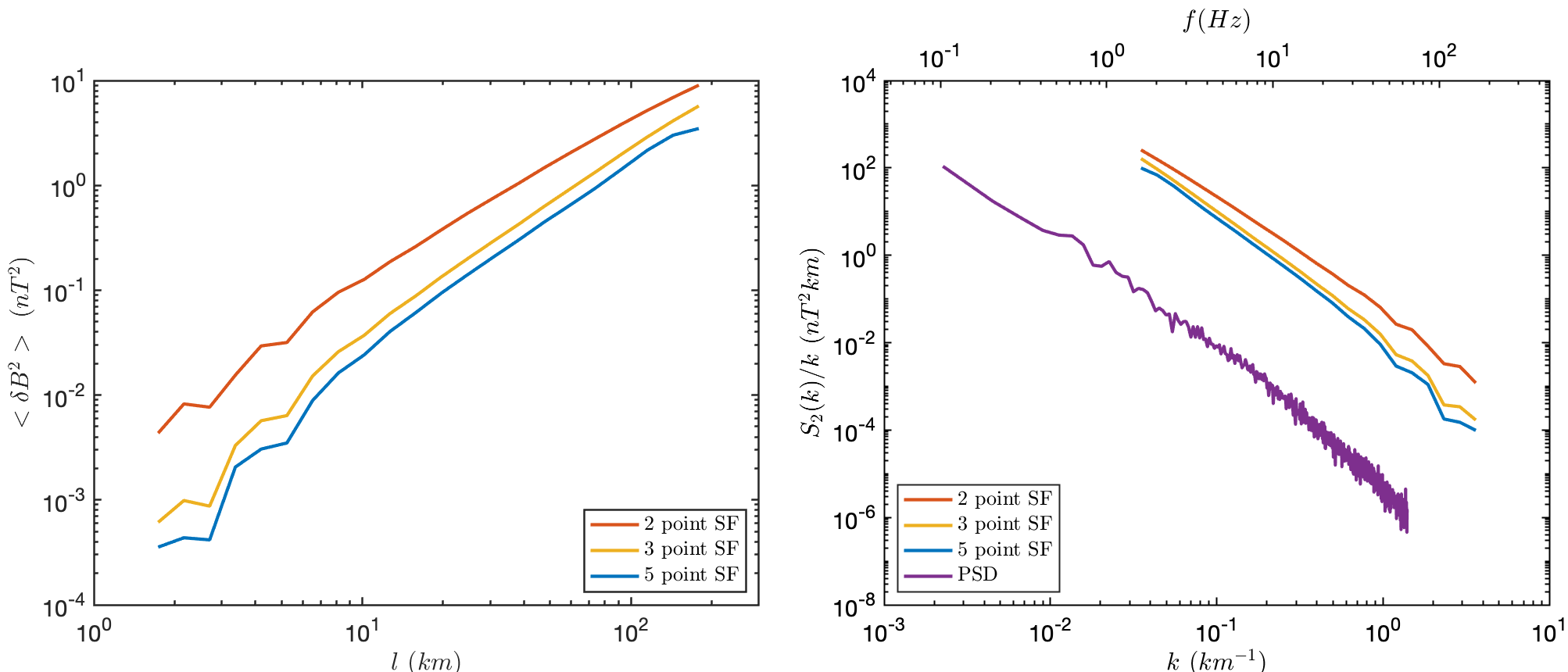}}
\caption{Comparison of multi-point structure functions during 08:02:13-08:12:33 on 4 Oct 2017.}
\label{fig:ap1}
\end{figure}

The differences between multi-point second-order structure functions of the total magnetic field are compared here, where the two-point and three-point structure functions are defined as $S_2^{(2)}(\bm{l};f)=\langle|f(\bm{r}+\bm{l})- f(\bm{r})|^2\rangle_{\bm{r}}$, and $S_2^{(3)}(\bm{l};f)=1/3 \langle|f(\bm{r}-\bm{l})- 2f(\bm{r})+ f(\bm{r}+\bm{l})|^2\rangle_{\bm{r}}$, respectively. As seen in the left panel of Figure \ref{fig:ap1}, the trend of three-point and five-point $SF$ tend to agree with each other, whereas the slope of two-point $SF$ are relatively flatter, especially towards small scales. We also compute the ``equivalent spectrum” defined as $S_2(k)/k$ and compare the results with power spectral density (PSD) from Fourier transform. Again, it is found that within [0.05, 1] $km^{-1}$ of the right panel, the slope of the spectrum based on five-point $SF$ is around -2.86, which is similar with the three-point result of -2.80 and the slope of PSD around -2.94, while the slope based on two-point $SF$ is only -2.55. Hence it is demonstrated that in order to capture the spectral characteristics of the turbulence at sub-ion regime, the use of multi-point ($>$2) structure functions are preferred. 

\section{Validity of the Taylor hypothesis}\label{app:a2}
At kinetic scales, the Taylor hypothesis may be violated by to the significant fluctuations in the turbulent flows, or due to the large phase speed of the fluctuations exceeding the bulk flow velocity (e.g. \citet{2019EP&S...71...41T, 2019ApJ...876..138H}). The validity of Taylor hypothesis for all the events is checked by comparing the structure function of magnetic fluctuations in two ways \citep{2017ApJ...842..122C}: one is to calculate the structure function from single-spacecraft measurements by assuming Taylor hypothesis, and the other is to calculate the structure function based on direct spatial differences between measurements from six pairs of MMS spacecrafts, which are separated by certain inter-distances between them.

\begin{figure}[ht!]
\centerline{\includegraphics[width=0.6\textwidth,trim={0cm 0cm 0cm 0cm},clip=]{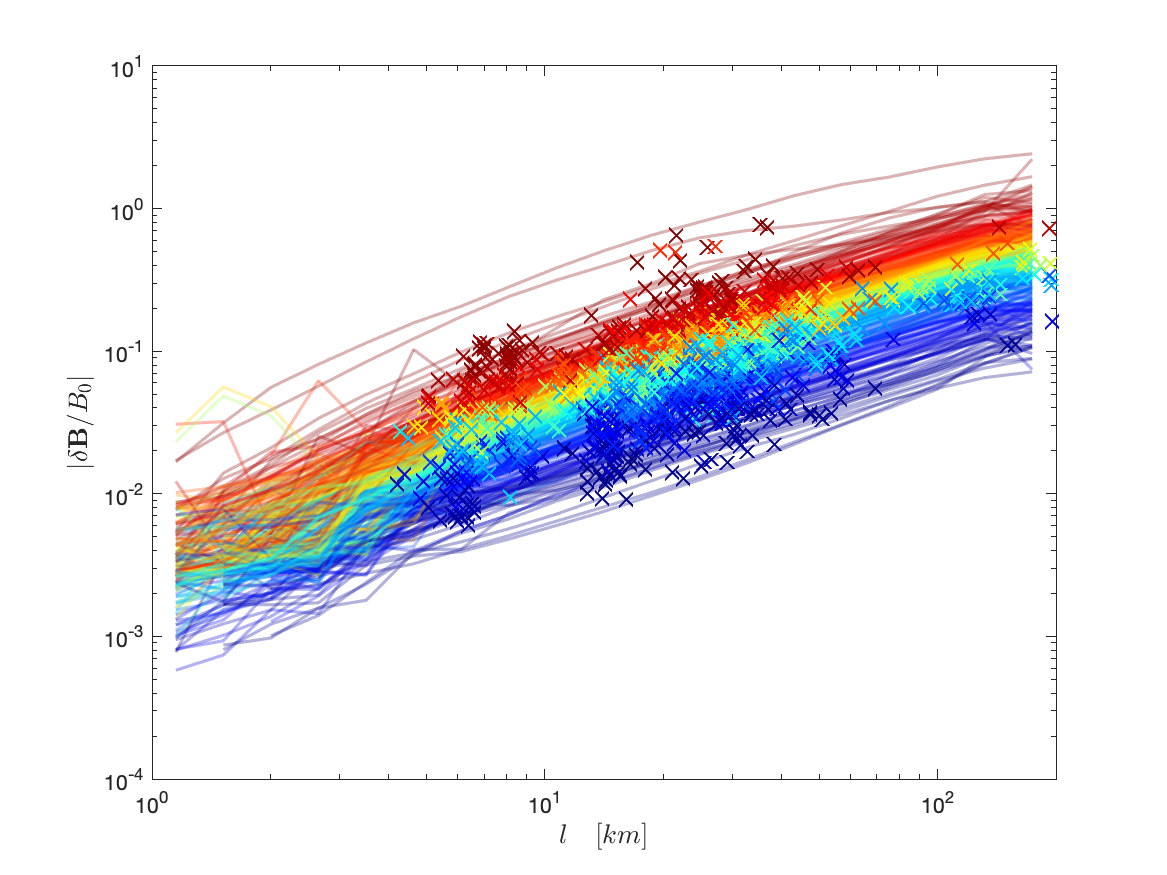}}
\caption{Test of validity of the Taylor hypothesis. The vertical axis represents the normalized magnetic fluctuation amplitudes obtained from first-order structure function. The curves are from the time differences of measurements from individual spacecraft under Taylor assumption, while the crosses are based on the spatial differences between measurements from the six pairs of MMS spacecraft.}
\label{fig:ap2}
\end{figure}

Figure \ref{fig:ap2} shows the statistical results of the first-order structure function as a function of scale. As represented by different colour for each individual event, the results based on Taylor hypothesis (curves) are close to the results from direct spatial measurements (crosses) at 5 km $< l <$ 200 km. Therefore, the use of Taylor hypothesis in our analysis have been proven to be reasonable. We note that our results at small scales are in agreement with recent demonstration of Taylor hypothesis being valid down to $k d_e \sim 1$ \citep{2017ApJ...842..122C}. In addition, the Taylor condition are known to be better satisfied at relatively large perpendicular wavenumbers especially when fluctuations are sampled along the perpendicular direction (\citet{2017ApJ...842..122C} and references therein), thus the presence of spectral anisotropy ($k_{\perp}>k_{\parallel}$) in our events is also in favour of the Taylor assumptions.

\bibliography{sample63}{}
\bibliographystyle{aasjournal}



\end{document}